\documentstyle[11pt]{article}

\let\a=\alpha \let\b=\beta \let\g=\gamma \let\d=\delta \let\e=\epsilon
    
 \let\m=\mu \let\n=\nu   
   \let\f=\phi  
       \let\D=\Delta

\def\nn{\nonumber} \def\bd{\begin{document}} \def\ed{\end{document}}
\def\ds{\documentstyle} \let\fr=\frac \let\bl=\bigl \let\br=\bigr
\let\Br=\Bigr \let\Bl=\Bigl 
\let\bm=\bibitem
\let\na=\nabla
\let\pa=\partial \let\ov=\overline 
\newcommand{\be}{\begin{equation}} 
\newcommand{\ee}{\end{equation}} 
\def\ba{\begin{array}}
\def\ea{\end{array}}
\def\ft#1#2{{\textstyle{{\scriptstyle #1}\over {\scriptstyle #2}}}}
\def\fft#1#2{{#1 \over #2}}
\def\del{\partial}
\def\vp{\varphi}
\def\sst#1{{\scriptscriptstyle #1}}
\def\oneone{\rlap 1\mkern4mu{\rm l}}
\def\td{\tilde}
\def\wtd{\widetilde}
\def\ie{\rm i.e.\ }
\def\dalemb#1#2{{\vbox{\hrule height .#2pt
        \hbox{\vrule width.#2pt height#1pt \kern#1pt
                \vrule width.#2pt}
        \hrule height.#2pt}}}
\def\square{\mathord{\dalemb{6.8}{7}\hbox{\hskip1pt}}}
\newcommand{\ho}[1]{$\, ^{#1}$}
\newcommand{\hoch}[1]{$\, ^{#1}$}
\newcommand{\bea}{\begin{eqnarray}} 
\newcommand{\eea}{\end{eqnarray}} 
\newcommand{\ra}{\rightarrow}
\newcommand{\lra}{\longrightarrow}
\newcommand{\Lra}{\Leftrightarrow}
\newcommand{\ap}{\alpha^\prime}
\newcommand{\bp}{\tilde \beta^\prime}
\newcommand{\tr}{{\rm tr} }
\newcommand{\Tr}{{\rm Tr} } 
\def\0{{\sst{(0)}}}
\def\1{{\sst{(1)}}}
\def\2{{\sst{(2)}}}
\def\3{{\sst{(3)}}}
\def\4{{\sst{(4)}}}
\def\5{{\sst{(5)}}}
\def\6{{\sst{(6)}}}
\def\7{{\sst{(7)}}}
\def\8{{\sst{(8)}}}
\def\n{{\sst{(n)}}}
\def\cA{{{\cal A}}}
\def\cF{{{\cal F}}}
\def\tV{\widetilde V}
\def\tW{\widetilde W}
\def\tH{\widetilde H}
\def\tE{\widetilde E}
\def\tF{\widetilde F}
\def\tA{\widetilde A}
\def\im{{{\rm i}}}
\def\tY{{{\wtd Y}}}
\def\ep{{\epsilon}}
\def\vep{{\varepsilon}}
\def\R{\rlap{\rm I}\mkern3mu{\rm R}}
\def\bD{{{\bar D}}}

\newcommand{\NP}{Nucl. Phys. }
\newcommand{\tamphys}{\it Center for Theoretical Physics,
Texas A\&M University, College Station, TX 77843}
\newcommand{\upenn}{\it Dept. of Phys. and Astro.,
University of Pennsylvania,
Philadelphia, PA 19104}

\newcommand{\auth}{M. Cveti\v{c}\hoch{\dagger1}, H. L\"u\hoch{\dagger1},
C.N. Pope\hoch{\ddagger2}, A. Sadrzadeh\hoch{\ddagger} and
  T.A. Tran\hoch{\ddagger}}

\begin{document}
\begin{flushright}
\hfill{CTP TAMU-13/00 \\
UPR/888-T \\
May 2000}\\
\hfill{\bf hep-th/0005137}\\
\end{flushright}

\vspace{10pt}

\begin{center}
{\large {\bf $S^3$ and $S^4$ Reductions of Type IIA Supergravity}}

\vspace{20pt}

\auth

\vspace{10pt}
{\hoch{\dagger}\upenn}

\vspace{10pt}
{\hoch{\ddagger}\tamphys}

\vspace{30pt}

\underline{ABSTRACT}
\end{center}

     We construct a consistent reduction of type IIA supergravity on
$S^3$, leading to a maximal gauged supergravity in seven dimensions
with the full set of massless $SO(4)$ Yang-Mills fields.  We do this
by starting with the known $S^4$ reduction of eleven-dimensional
supergravity, and showing that it is possible to take a singular limit
of the resulting standard $SO(5)$-gauged maximal supergravity in seven
dimensions, whose eleven-dimensional interpretation involves taking a
limit where the internal 4-sphere degenerates to $\R\times S^3$.  This
allows us to reinterpret the limiting $SO(4)$-gauged theory in seven
dimensions as the $S^3$ reduction of type IIA supergravity.  We also
obtain the consistent $S^4$ reduction of type IIA supergravity, which
gives an $SO(5)$-gauged maximal supergravity in $D=6$.

{\vfill\leftline{}\vfill
\vskip 10pt \footnoterule {\footnotesize \hoch{1} Research supported
in part by DOE grant DOE-FG02-95ER40893
\vskip  -12pt} \vskip   14pt
{\footnotesize
        \hoch{2}        Research supported in part by DOE
grant DOE-FG03-95ER40917 \vskip -12pt}  \vskip  14pt
}

\pagebreak
\setcounter{page}{1}

\section{Introduction}

         The study of Kaluza-Klein sphere reductions of supergravities
has so far concentrated mostly on the examples where the theories
admit vacuum solutions of the form AdS$\times$Sphere, which are the
near-horizon structures of certain $p$-brane solutions of the
theories.\footnote{Throughout this paper we are concerned only with
those ``remarkable'' Kaluza-Klein sphere reductions for which no known
group-theoretic argument guaranteeing the consistency of the reduction
exists.  Consistent reductions on $S^3$, or indeed any group manifold
$G$, can always be performed in the case where one truncates to the sector
of singlets under the right action of the group $G$, but the
consistency in such a case is guaranteed, and therefore is not of
interest to us in the present paper.} 
  These include 11-dimensional supergravity, which has
AdS$_4\times S^7$ and AdS$_7\times S^4$ vacuum solutions, and type IIB
supergravity, which has an AdS$_5\times S^5$ vacuum solution.  The
$S^7$ and $S^4$ reduction An\"atze for 11-dimensional supergravity
were presented in \cite{dwn,vann2}.  For type IIB supergravity the
$S^5$ reduction of its $SL(2,\R)$-singlet subsector, which gives rise
to five-dimensional fields including the entire set of $SO(6)$ gauge
bosons, was given in \cite{clpst}.  Explicit reduction Ans\"atze for
various subsectors of these supergravity reductions can be found in
\cite{ten,d7gauge,d5gauge,d4gauge,dist,clps,clpd45,pw,BS2}.

         In general, vacuum supergravity solutions with non-trivial
field-strength fluxes are of the form of warped products of a certain
spacetime geometry and internal spheres.  The consistency of sphere
reductions in such cases have been much less fully studied.  The first
example of this type was the consistent warped $S^4$ reduction
\cite{d6gauge} of massive type IIA supergravity, to give rise to the
massive $SU(2)$-gauged supergravity in $D=6$.  The vacuum AdS$_6$
solution can be viewed as the near-horizon structure \cite{oz} of an
intersecting D4-D8 brane \cite{youm}.

          Further examples of consistent sphere reductions were
obtained in \cite{s3red}, where the resulting theories admit ``vacuum
solutions'' that are domain walls rather than AdS spacetimes.  In
\cite{s3red}, a necessary condition for the consistency of a sphere
reduction of a theory was given.  Namely, if a theory can be
consistently reduced on $S^n$, with a massless truncation that retains
all the $SO(n+1)$ Yang-Mills gauge fields, then a necessary
requirement is that if instead a toroidal reduction on $T^n$ is
performed, this must give rise to the same content of massless
fields.\footnote{This is because by turning off the gauge-coupling
parameter $g$, by sending the radius of the $n$-sphere to infinity, we
must recover the same massless field content as would result from a
flat (toroidal) reduction.}  Furthermore, the $T^n$-reduced theory
must have at least an $SO(n+1)$ global symmetry, with sufficiently
many abelian vector fields to supply at least those of the adjoint
representation of $SO(n+1)$.  These conditions are very restrictive,
and only in limited cases can a consistent sphere reduction that
retains all the Yang-Mills fields occur.

     In the type IIA and type IIB theories, the NS-NS branes and
D-branes have near-horizon structures of the form (Domain wall)$\times
S^n$, for various values of $n$.  It is easy to verify in each case
that if a Kaluza-Klein reduction of the theory on $T^n$ is performed,
this yields a scalar coset $SL(n+1,\R)/SO(n+1)$ in the lower
dimension, since this theory can instead be obtained from a $T^{n+1}$
reduction from $D=11$.  In particular, therefore, this theory has an
$SO(n+1)$ global symmetry subgroup.  It was noted \cite{bst} that
there exists a ``dual frame'' for each D-brane, in which the
near-horizon structure of the brane generically becomes
AdS$_{10-n}\times S^n$, while instead it becomes Minkowski$\times S^3$
when $n=3$.  This leads to the conjectured Domain Wall/QFT
correspondence \cite{bst}, generalising the notion of the AdS/CFT
correspondence \cite{malda,gkp,wit}.  This correspondence was
generalised to lower dimensions in \cite{bh}.  

        Further evidence for Domain Wall/QFT correspondence was
obtained in \cite{distdw}, where it was shown that it is consistent to
reduce to the subsector scalars associated with the Cartan generators
in the scalar coset $SL(n+1,\R)/SO(n+1)$.  The multi-parameter
domain-wall solutions of these lower-dimensional theories can then be
lifted back to the higher dimension, where they correspond to certain
ellipsoidal distributions of the $p$-branes, thus implying that these
domain wall geometries correspond to the Coulomb branch of the quantum
field theory, and generalising the results on the Coulomb branch in
the AdS/CFT correspondence, discussed in
\cite{KLT,FGPW,BS,BSI,dist,BBS}. Interestingly, the wave equations for
minimally-coupled scalar fluctuations in the lower-dimensional
domain-wall backgrounds depend only on the dimension of the internal
sphere used in the reduction \cite{distdw}.  

       The consistency of the Kaluza-Klein reduction in this Cartan
subsector of scalar fields leads one to believe further that it is
consistent to reduce the type IIA and type IIB theories on the
relevant $n$-spheres, while retaining all the massless fields.  For
non-trivial vacuum NS-NS flux, both the type IIA and type IIB theories
can be expected to be consistently reducible on $S^3$ and on $S^7$,
and indeed, for $N=1$ supergravity, the consistency has been
demonstrated, and the corresponding gauged supergravities in $D=7$ and
$D=3$ were obtained in \cite{s3red}.  For non-trivial vacuum R-R flux,
we expect that it is consistent to reduce the type IIA theory on $S^n$
with $n=2,4,6,8$ and the type IIB theory on $S^n$ with
$n=1,3,5,7$. Indeed, the $S^4$ and $S^7$ reductions of type IIA, where
$S^4$ and $S^7$ are associated with the D4-brane and NS-NS string, can
be established from the $S^1$ reduction of the corresponding $S^4$ or
$S^7$ reduction of eleven-dimensional supergravity.  In section 5, we
carry this out explicitly for the $S^4$ reduction of the type IIA theory.

      First, we demonstrate explicitly that it is consistent
to perform an $S^3$ reduction of the type IIA theory, while retaining
all the massless fields, including in particular the entire set of
$SO(4)$ Yang-Mills gauge fields.  This case is of particular interest
because $S^3$ is itself the group manifold $SU(2)$, and strings
propagating in group-manifold backgrounds have been extensively
studied in the past.  It should be emphasised though that typically
when Kaluza-Klein reductions on a group manifold $G$ have been
discussed in the literature, a truncation is performed in which only
those fields that are singlets under the left action $G_L$ of the
$G_L\times G_R$ isometry group are retained.  For example, the $S^3$
reduction of $N=1$ supergravity in $D=10$, retaining only one $SU(2)$
Yang-Mills fields, was performed to give rise to gauged simple
supergravity in $D=7$ with a domain wall vacuum solution
\cite{chamsab}. Such a truncation guarantees that a consistent
reduction can be performed, but it fails to exploit the much more
remarkable fact that in this $S^3$ case a reduction that retains {\it
all} the $SO(4)\sim SU(2)_L\times SU(2)_R$ Yang-Mills fields, and not
merely those of $SU(2)_L$, is possible.  A further reason for wishing
to include all the gauge fields of $SO(4)$ is that only then do we
obtain a seven-dimensional theory with maximal supersymmetry.

    We obtain the consistent $S^3$ reduction of type
IIA supergravity by taking a singular limit of the $S^4$ reduction of
eleven-dimensional supergravity, in which the $S^4$ degenerates to
$\R\times S^3$.  In order to do this, we begin in section 2 by
reviewing the $S^4$ reduction from $D=11$, first obtained in
\cite{vann2}.  By substituting this into the eleven-dimensional
Bianchi identity and equation of motion for the 4-form, we obtain
complete and explicit seven-dimensional equations of motion, and the
Lagrangian that generates them.  We also discuss the ``ungauging
limit'' in which the radius of the 4-sphere is sent to infinity.  (An
analogous limit was also considered in \cite{jim}, in the context of
the $U(1)^2$ subgroup of $SO(4)$.)  In particular, we clarify certain
aspects of this limiting process, showing that the limit is smooth in
the seven-dimensional equations of motion, but pathological at the
level of the gauged-supergravity Lagrangian.

    In section 3 we take a different singular limit of the
seven-dimensional $SO(5)$-gauged supergravity, in which an $SO(4)$
gauging remains.  Again, this is a smooth limit of the equations of
motion, but not of the gauged supergravity Lagrangian.  In section 4
we apply this limiting procedure to the $S^4$ Kaluza-Klein reduction
Ansatz of eleven-dimensional supergravity, showing that it corresponds
to a degeneration of the 4-sphere to $\R\times S^3$.  The reduction
can then be viewed as an initial reduction to give type IIA
supergravity in $D=10$, followed by a reduction on $S^3$.  By this
means, we arrive at the consistent $S^3$ reduction Ansatz for type IIA
supergravity.

    In section 5 we construct instead the consistent $S^4$
Kaluza-Klein reduction of type IIA supergravity.  This can again be
obtained from the starting point of the $S^4$ reduction of the
eleven-dimensional theory.  In this case we do not need to take any
singular limit in the internal directions, but rather, we perform a
standard Kaluza-Klein $S^1$ reduction of the original
seven-dimensional theory coming from $D=11$, and show how this can be
reinterpreted as an $S^4$ reduction of type IIA supergravity.  The
paper ends with concluding remarks in section 6.

\section{The $S^4$ reduction of eleven-dimensional supergravity}

\subsection{Metric and 4-form Ansatz}

   The complete Ansatz for the $S^4$ reduction of eleven-dimensional
supergravity was obtained in \cite{vann2}, using a formalism based on
an analysis of the supersymmetry transformation rules.  One may also
study the reduction from a purely bosonic standpoint, by verifying
that if the Ansatz is substituted into the eleven-dimensional
equations of motion, it consistently yields the equations of motion of
the seven-dimensional gauged $SO(5)$ supergravity.  We shall carry out
this procedure here, in order to establish notation, and to obtain the
complete system of seven-dimensional bosonic equations of motion,
which we shall need in the later part of the paper.

     After some manipulation, the Kaluza-Klein $S^4$ reduction Ansatz
obtained in \cite{vann2} for eleven-dimensional supergravity can be
expressed as follows:
\bea
d\hat s_{11}^2 &=& \Delta^{1/3}\, ds_{7}^2 + \fr1{g^2}\Delta^{-2/3}\,
T^{-1}_{ij}\, D\mu^i\, D\mu^j\,,\label{metel} 
\eea
\bea
\hat F_\4 &=& \fft1{4!}\, \ep_{i_1\cdots i_5}\, \Big[
- \fr1{g^3} U\, \Delta^{-2} \mu^{i_1} D\mu^{i_2}\wedge \cdots \wedge
D\mu^{i_5}\nn\\
&& + \fr4{g^3} \Delta^{-2}\, T^{i_1 m}\, DT^{i_2 n}\, \mu^m\, \mu^n\,
D\mu^{i_3}
\wedge \cdots \wedge D\mu^{i_5}\nn\\
&& + \fr6{g^2} \Delta^{-1} F_\2^{i_1 i_2} \wedge
D\mu^{i_3}\wedge D\mu^{i_4}\, T^{i_5 j}\, \mu^j \Big] - T_{ij}\,
{*S_\3^i}\, \mu^j + \fft1{g}\, S_\3^i \wedge D\mu^i\,,\label{4form}
\eea
where  
\bea
U \equiv 2 T_{ij}\, T_{jk}\, \mu^i\, \mu^k - \Delta\, T_{ii}\,, \qquad
\Delta \equiv T_{ij}\, \mu^i\, \mu^j\,,\nn\\
F_\2^{ij} \equiv dA_\1^{ij} + g A_\1^{ik}\wedge A_\1^{kj}\,,
\qquad  D\mu^i \equiv d\mu^i + g A_\1^{ij}\, \mu^j\,,\nn\\
DT_{ij} \equiv dT_{ij} + g A_\1^{ik}\, T_{kj} + g A_\1^{jk}\, T_{ik}\,,
\qquad \mu^i\, \mu^i \equiv 1\,,
\eea
where the symmetric matrix $T_{ij}$, which parameterises the scalar
coset $SL(6,\R)/SO(6)$, is unimodular.
 
\subsection{Derivation of the seven-dimensional equations of motion}

    Consider first the Bianchi identity $d\hat F_\4 = 0$.
Substituting (\ref{4form}) into this, we obtain the following
seven-dimensional equations:
\bea
D(T_{ij}\, {* S_\3^j}) &=& F_\2^{ij}\wedge S_\3^j\,,\label{bianchiv}\\
H_\4^i &=& g T_{ij}\, {* S_\3^j} +  \fft1{8}  \ep_{i
{j_1}\cdots {j_4}} F_\2^{{j_1} {j_2}}\wedge\, F_\2^{{j_3} {j_4}}\,,
\label{h4eq}
\eea
where we define
\be
H_\4^i \equiv D S_\3^i = dS_\3^i + g\, A_\1^{ij}\wedge S_\3^j\,.
\label{h4def}
\ee

    Next, we substitute the Ansatz into the $D=11$ field equation 
$d {{\hat *}\hat F_\4} = \ft12 {\hat F_\4}\wedge{\hat F_\4}$.  To do
this, we need the eleven-dimensional Hodge dual ${\hat *\hat F_\4}$,
which we find is given by
\bea
{{\hat *}\hat F_\4} &=& - g U \ep_\7 - \fr1{g} T^{-1}_{ij} {*D}T^{ik}
\mu_k\wedge D\mu^j +\fr1{2g^2} T^{-1}_{ik} T^{-1}_{j\ell}\, 
{*F_\2^{ij}} \wedge D\mu^k \wedge D\mu^\ell \\
&& \!\!\!\!\!
+ \fr1{g^4} \Delta^{-1}\, T_{ij}\, S_\3^i\, \mu^j \wedge W - \fft1{6 g^3}
\, \Delta^{-1}\, 
\ep_{ij \ell_1\ell_2 \ell_3} {*S_\3^m}\, T_{im}\,  T_{jk}\, \mu^k 
\wedge D\mu^{\ell_1}\wedge D\mu^{\ell_2} \wedge D\mu^{\ell_3}\,,\nn
\eea
where
\be
W\equiv \ft1{24}\, \ep_{i_1\cdots i_5}\, \mu^{i_1}\, D\mu^{i_2}\wedge 
\cdots \wedge D\mu^{i_5}\,.
\ee
The field equation for $\hat F_\4$ then implies
\bea
{D\Big(T^{-1}_{ik} T^{-1}_{j\ell} {*F_\2^{ij}}\Big)} &=& -2 g\, 
T^{-1}_{i[k} {*DT_{\ell] i}} 
- \fft1{2g}\,
\ep_{i_1 i_2 i_3 k \ell}\, F_2^{i_1 i_2}\, H_\4^{i_3}
\nn\\
&&
+ \fr3{2g} \delta_{i_1 i_2 k\ell}^{j_1 j_2 j_3 j_4}\, F_\2^{i_1 i_2}\wedge
F_\2^{j_1 j_2}\wedge  F_\2^{j_3 j_4} -
 S_\3^k\wedge S_\3^\ell\,.\label{gaugev}\\
D\Big(\, T^{-1}_{ik} {*D(T_{kj}})\Big) &=& 2 g^2 (2 T_{ik}\, T_{kj} -
T_{kk}\,
T_{ij})\ep_\7 + T^{-1}_{im}\, T^{-1}_{k\ell}\, 
{*F_\2^{m\ell}}\wedge F_\2^{kj}\nn\\ 
&& + T_{jk}\, {*S_\3^k} \wedge S_\3^i - \ft15 \delta_{ij}
\Big[ 2 g^2 \Big(2T_{ik} T_{ik} - 2 (T_{ii})^2 \Big) \ep_\7  \nn\\
&& + T^{-1}_{nm} T^{-1}_{k\ell}\, {*F_\2^{m\ell}} \wedge F_\2^{kn} + 
T_{k\ell } \, {*S_\3^k} \wedge S_\3^\ell \Big]\,, \label{scalarsv}
\eea
for the Yang-Mills and scalar equations of motion in
$D=7$.\footnote{Note from (\ref{gaugev}) that it would be inconsistent
to set the Yang-Mills fields to zero while retaining the scalars
$T_{ij}$, since the currents $T^{-1}_{i[k} {*DT_{\ell] i}}$ act as
sources for them.  A truncation where the Yang-Mills fields are set to
zero {\it is} consistent, however, if the scalars are also truncated
to the diagonal subsector $T_{ij}={\rm diag}(X_1,X_2,\ldots, X_6)$, as
in the consistent reductions constructed in \cite{dist,clps}.}

     We find that all the  equations of motion can be derived from the
following seven-dimensional Lagrangian
\bea
{\cal L}_7 &=& R\, {*\oneone} - 
\ft14 T^{-1}_{ij}\, {*D T_{jk}}\wedge
T^{-1}_{k\ell}\, D T_{\ell i}  
-\ft1{4}\, T^{-1}_{ik}\, T^{-1}_{j\ell}\, {* F_\2^{ij}}\wedge F_\2^{k\ell}
-\ft12 T_{ij}\, {*S_\3^i}\wedge S_\3^j \nn\\
&&+ \fft1{2g} S_\3^i\wedge H_\4^i -  
\fft1{8g}  \ep_{i j_1\cdots j_4}\, S_\3^i\wedge F_\2^{j_1 j_2}\wedge 
F_\2^{j_3 j_4} + \fr1g \Omega_\7 - V\, {*\oneone}\,,\label{d7lag}
\eea
where $H_\4^i$ are given by (\ref{h4def}) and the potential $V$ is
given by
\be
V = \ft12  g^2 \Big(2 T_{ij}\, T_{ij} - (T_{ii})^2 \Big)\,,
\ee
and $\Omega_\7$ is a Chern-Simons type of term built from the
Yang-Mills fields, which has the property that its variation with
respect to $A_\1^{ij}$ gives
\be
\delta \Omega_\7 =  
\ft34 \delta_{i_1 i_2 k\ell}^{j_1 j_2 j_3 j_4}\, F_\2^{i_1 i_2}\wedge
F_\2^{j_1 j_2}\wedge  F_\2^{j_3 j_4}\wedge \delta A_\1^{k\ell}\,.
\ee
Note that the $S_\3^i$ are viewed as fundamental fields in the Lagrangian,
and that (\ref{h4eq}) is their first-order equation.  In fact
(\ref{d7lag}) is precisely the bosonic sector of the Lagrangian
describing maximal gauged seven-dimensional supergravity that was
derived in \cite{ppn}.  An explicit expression for the 7-form
$\Omega_\7$ can be found there.

    Although we have fully checked the eleven-dimensional Bianchi
identity and field equation for $\hat F_\4$ here, we have not
completed the task of substituting the Ansatz into the
eleven-dimensional Einstein equations.  This would be an extremely
complicated calculation, on account of the Yang-Mills gauge fields.
However, various complete consistency checks, including the
higher-dimensional Einstein equation, have been performed in various
truncations of the full $N=4$ maximal supergravity embedding,
including the $N=2$ gauged theory in \cite{d7gauge}, and the
non-supersymmetric truncation in \cite{clps} where the gauge fields
are set to zero and only the diagonal scalars in $T_{ij}$ are
retained.  All the evidence points to the full consistency of the
reduction.\footnote{The original demonstration in \cite{vann2}, based
on the reduction of the eleven-dimensional supersymmetry
transformation rules, also provides extremely compelling evidence.
Strictly speaking the arguments presented there also fall short of a
complete and rigorous proof, since they involve an approximation in
which the quartic fermion terms in the theory are neglected.}

    It is perhaps worth making a few further remarks on the nature of
the reduction Ansatz.  One might wonder whether the Ansatz
(\ref{4form}) on the 4-form field strength $\hat F_\4$ could be
re-expressed as an Ansatz on its potential $\hat A_\3$.  As it stands,
(\ref{4form}) only satisfies the Bianchi identity $d\hat F_\4=0$ by
virtue of the lower-dimensional equations (\ref{bianchiv}) and
(\ref{h4eq}).  However, if (\ref{h4eq}) is substituted into
(\ref{4form}), we obtain an expression that satisfies $d\hat F_\4=0$
without the use of any lower-dimensional equations.  However, one does
still have to make use of the fact that the $\mu^i$ coordinates
satisfy the constraint $\mu^i\, \mu^i=1$, and this prevents one from
writing an explicit Ansatz for $\hat A_\3$ that has a manifest $SO(5)$
symmetry.  One could solve for one of the $\mu^i$ in terms of the
others, but this would break the {\it manifest} local symmetry from
$SO(5)$ to $SO(4)$.  In principle though, this could be done, and then
one could presumably substitute the resulting Ansatz directly into the
eleven-dimensional Lagrangian.  After integrating out the internal
4-sphere directions, one could then in principle obtain a
seven-dimensional Lagrangian in which, after re-organising terms, the
local $SO(5)$ symmetry could again become manifest.

    It should, of course, be emphasised that merely substituting an
Ansatz into a Lagrangian and integrating out the internal directions
to obtain a lower-dimensional Lagrangian is justifiable only if one
already has an independent proof of the consistency of the proposed
reduction Ansatz.\footnote{A classic illustration is provided by the
example of 5-dimensional pure gravity with an (inconsistent)
Kaluza-Klein reduction in which the scalar dilaton is omitted.
Substituting this into the 5-dimensional Einstein-Hilbert action
yields the perfectly self-consistent Einstein-Maxwell action in $D=4$,
but fails to reveal that setting the scalar to zero is inconsistent
with the internal component of the 5-dimensional Einstein equation.}
If one is in any case going to work with the higher-dimensional field
equations in order to prove the consistency, it is not clear that
there would be any significant benefit to be derived from then
re-expressing the Ansatz in a form where it could be substituted into
the Lagrangian.

\subsection{Ungauging: the $g\rightarrow 0$ limit}

   It is interesting to observe that one cannot take the limit
$g\rightarrow0$ in the Lagrangian (\ref{d7lag}), on account of the
terms proportional to $g^{-1}$ in the second line.  We know, on the
other hand, that it must be possible to recover the ungauged $D=7$
theory by turning off the gauge coupling constant.  In fact the
problem is associated with a pathology in taking the limit at the
level of the Lagrangian, rather than in the equations of motion.  This
can be seen by looking instead at the seven-dimensional equations of
motion, which were given earlier.  The only apparent obstacle to
taking the limit $g\rightarrow0$ is in the Yang-Mills equations
(\ref{gaugev}), but in fact this illusory.  If we substitute the
first-order equation (\ref{h4eq}) into (\ref{gaugev}) it gives
\be
{D\Big(T^{-1}_{ik} T^{-1}_{j\ell} {*F_\2^{ij}}\Big)} = -2 g T^{-1}_{i[k}
{*DT_{\ell] i}} - \ft1{2}\,
\ep_{i_1 i_2 i_3 k \ell}\, F_2^{i_1 i_2}\wedge T_{ij}\, {*S_\3^j}
-S_\3^k\wedge S_\3^\ell\,,
\label{gaugev2}
\ee
which has a perfectly smooth $g\rightarrow 0$ limit.  It is clear
that equations of motion (\ref{h4eq}) and (\ref{scalarsv}) and the
Einstein equations of motion also have a smooth limit.  (The
reason why the Einstein equations have the smooth limit is because the
$1/g$ terms in the Lagrangian (\ref{d7lag}) do not involve the metric,
and thus they give no contribution.)

    Unlike in the gauged theory, we should not treat the $S^i_\3$
fields as fundamental variables in a Lagrangian formulation in the
ungauged limit.  This is because once the gauge coupling $g$ is sent
to zero, the fields $S^i_\3$ behave like 3-form field strengths.  This
can be seen from the first-order equation of motion (\ref{h4eq}),
which in the limit $g\rightarrow 0$ becomes
\be
dS_\3^i = \ft18 \ep_{ij_1\cdots j_4}\, dA_\1^{j_1j_2}\wedge 
dA_\1^{j_3j_4}\,,\label{s3ibianchi}
\ee
and should now be interpreted as a Bianchi identity.  This can be
solved by introducing 2-form gauge potentials $A_\2^i$, with the
$S_\3^i$ given by
\be
S_\3^i = dA_\2^i + \ft18 \ep_{ij_1\cdots j_4}\, A_\1^{j_1j_2}\wedge 
dA_\1^{j_3j_4}\,.\label{s3isol}
\ee
In terms of these 2-form potentials, the equations of motion can
now be obtained from the Lagrangian
\bea
{\cal L}^0_7 &=& R\, {*\oneone} - 
\ft14 T^{-1}_{ij}\, {*d T_{jk}}\wedge
T^{-1}_{k\ell}\, d T_{\ell i}  
-\ft1{4}\, T^{-1}_{ik}\, T^{-1}_{j\ell}\, {* F_\2^{ij}}\wedge
F_\2^{k\ell}-\ft12 T_{ij}\, {*S_\3^i}\wedge S_\3^j\nn\\
&&+ \ft12 A_\1^{ij}\wedge S_\3^i\wedge S_\3^j -2 S_\3^i\wedge A_\2^j
\wedge dA_\1^{ij}
\,,\label{d7g0lag} 
\eea
where $S_\3^i$ is given by (\ref{s3isol}).  This is precisely the
bosonic Lagrangian of the ungauged maximal supergravity in $D=7$.

       It is worth exploring in a little more detail why it is
possible to take a smooth $g\rightarrow 0$ limit in the
seven-dimensional equations of motion, but not in the Lagrangian.  We
note that in this limit the Lagrangian (\ref{d7lag}) can be expressed
as
\be
{\cal L}_7 = \fft1{g}\, L + {\cal O}(1)\,,
\ee
where
\be
L = \ft12 S_\3^i dS_\3^i -\ft18 \ep_{i j_1\cdots j_4}\, 
S_\3^i\wedge F_\2^{j_1 j_2}\wedge 
F_\2^{j_3 j_4} + \Omega_\7\,.\label{Lterms}
\ee       
The term $L/g$, which diverges in the $g\rightarrow 0$ limit, clearly
emphasises that the Lagrangian (\ref{d7g0lag}) is not merely the
$g\rightarrow 0$ limit of (\ref{d7lag}).  However if we make use of
the equations of motion, we find that in the $g\rightarrow 0$ limit
the $S_\3^i$ can be solved by (\ref{s3isol}).  Substituting this into
(\ref{Lterms}), we find that in this limit it becomes
\be
L=\ft1{16} \epsilon_{ij_1\cdots j_4} dA_\2^i\wedge dA_\1^{j_1j_2}\wedge
dA_\1^{j_3j_4} + {\cal O}(g)\,,
\ee
and so the singular terms in $L/g$ form a total derivative and hence
can be subtracted from the Lagrangian.  This analysis explains why it
is possible to take a smooth $g\rightarrow 0$ limit in the equations
of motion, but not in the Lagrangian.

\section{The gauged $SO(4)$ limit of maximal $D=7$ supergravity}

    Here we examine, at the level of the seven-dimensional theory
itself, how to take a limit in which the $SO(5)$ gauged sector is
broken down to $SO(4)$.  In a later section, we shall show how this
can be interpreted as an $S^3$ reduction of type IIA supergravity.  We
shall do that by showing how to take a limit in which the internal
$S^4$ in the original reduction from $D=11$ becomes $\R\times S^3$.
For now, however, we shall examine the $SO(4)$-gauged limit entirely
from the perspective of the seven-dimensional theory itself.

    To take the limit, we break the $SO(5)$ covariance by splitting
the $\underline 5$ index $i$ as
\be
i = (0,\a)\,,
\ee
where $1\le \a\le 4$.  We also introduce a constant parameter
$\lambda$, which will be sent to zero as the limit is taken.  We 
find that the various seven-dimensional fields, and the $SO(5)$ 
gauge-coupling  constant, should be scaled as follows:
\bea
&&g= \lambda^2\, \td g\,,\qquad A_\1^{0\a} = \lambda^3\, \wtd
A_\1^{0\a}\,,\qquad A_\1^{\a\b} = \lambda^{-2}\, \wtd
A_\1^{\a\b}\,,\nn\\
&&S_\3^0 = \lambda^{-4}\, \wtd S_\3^0\,,\qquad
S_\3^\a = \lambda\, \wtd S_\3^\a\,.\label{s3scal}\\
&&\nn\\
&&T^{-1}_{ij} = \pmatrix{\lambda^{-8}\, \Phi & \lambda^{-3}\, 
            \Phi\, \chi_\a\cr
            \lambda^{-3}\, \Phi\, \chi_\a & \lambda^2\, 
M^{-1}_{\a\b} + \lambda^2\,
\Phi\, \chi_\a\, \chi_\beta }\,.\nn
\eea
As we show in the next section, this rescaling corresponds to a degeneration
of $S^4$ to $R\times S^3$. Note that in this rescaling, we have also
performed a decomposition of the scalar matrix $T^{-1}_{ij}$ that is
of the form of a Kaluza-Klein metric decomposition. It is useful also
to present the consequent decomposition for $T_{ij}$, which turns out
to be
\be
T_{ij} =\pmatrix{ \lambda^8\, \Phi^{-1} + \lambda^8\, \chi_\gamma\,
\chi^\gamma &  - \lambda^3\, \chi^\a \cr
 -\lambda^3\, \chi^\a & \lambda^{-2}\, M_{\a\beta} }\,.
\ee
Calculating the determinant, we get
\be
\det(T_{ij}) = \Phi^{-1}\, \det(M_{\a\beta}) \,.
\ee
Since we know that $\det(T_{ij})=1$, it follows that
\be
\Phi = \det(M_{\a\beta})\,.\label{phim}
\ee
The fields $\chi_\a$ are ``axionic'' scalars.  Note that we shall also have
\bea
&&H_\4^0 = \lambda^{-4}\, \wtd H_\4^0\,,\qquad H_\4^\a = \lambda\,
\wtd H_\4^\a\,,\nn\\
&&\wtd H_\4^0 = d\wtd S_\3^0\,,\qquad 
\wtd H_\4^\a = \wtd D \wtd S_\3^\a - \td g\, \wtd A_\1^{0\a}\wedge
\wtd S_\3^0\,.
\eea
We have defined an $SO(4)$-covariant exterior derivative $\wtd D$,
which acts on quantities with $SO(4)$ indices $\a,\beta,\ldots$ in
the obvious way:
\be
\wtd D\, X_\a = d X_\a + \td g\, \wtd A_\1^{a\beta}\, X_\beta\,,
\ee
{\it etc}. 
It is helpful also to make the following further
field redefinitions:
\bea
G_\2^\a &\equiv& \wtd F_\2^{0\a} + \chi_\beta\, \wtd
F_\2^{\beta\a}\,,\nn\\
G_\3^\a &\equiv& \wtd S_\3^\a - \chi_\a\, \wtd S_\3^0\,,
\label{fieldredefs}\\
G_\1^\a &\equiv& \wtd D\chi_\a - \td g\, \wtd A_\1^{0\a}\,,\nn
\eea
where $\wtd F_\2^{0\a}\equiv \wtd D \wtd A_\1^{0\a}$.

    We may now subsititute these redefined fields into the
seven-dimensional equations of motion.  We find that a smooth limit in
which $\lambda$ is sent to zero exists, leading to an $SO(4)$-gauged 
seven-dimensional theory.  Our results for the seven-dimensional
equations of motion are as follows.  The fields $H_\4^i$ become
\be
\wtd H_\4^0 = d\wtd S_\3^0\,,\qquad \wtd H_\4^\a = \wtd D G_\3^\a +
G_\1^\a\wedge \wtd S_\3^0  +\chi_\a\, d S_\3^0\,.\label{h0ha}
\ee
The first-order equations (\ref{h4def}) give
\bea
\wtd H_\4^0 &=& \ft18 \ep_{\a_1\cdots \a_4}\, \wtd F_\2^{\a_1\a_2}\wedge 
                     \wtd F_\2^{\a_3\a_4}\,,\nn\\
\wtd F_\4^\a &=& \td g\, M_{\a\beta}\, {*G_\3^\beta} -\ft12 
\ep_{\a\beta\gamma\delta}\,G_\2^{\beta}\wedge \wtd
F_\2^{\gamma\delta} - G_\1^\a\wedge \wtd S_\3^0\,,
\label{s3firstorder}
\eea
where we have defined
\be
F_\4^\a\equiv \wtd D G_\3^\a\,.
\ee

   The second-order equations (\ref{bianchiv}) (which are nothing but
Bianchi identities following from (\ref{h4def})) become
\bea
d(\Phi^{-1}\, {*\wtd S_\3^0}) &=& M_{\a\beta}\, {*G_\3^\a}\wedge G_\1^\beta
            + G_\2^\a\wedge G_\3^\a\,,\nn\\
\wtd D(M_{\a\beta}\, {*G_\3^\beta}) &=& \wtd F_\2^{\a\beta}\wedge
G_\3^\beta - G_\2^\a\wedge \wtd S_\3^0\,.
\eea

    The Yang-Mills equations (\ref{gaugev}) become
\bea
\wtd D(\Phi\, M^{-1}_{\alpha\beta}\, {*G_\2^\beta}) &=&
\td g\, \Phi\, M_{\a\beta}\, {*G_\1^\beta}  - \wtd S_\3^0\wedge
G_\3^\a
- \ft12 \ep_{\a\beta_1\beta_2\beta_3}\, M_{\beta_3\gamma}\, \wtd
F_\2^{\beta_1\beta_2}\wedge {*G_\3^\gamma} \,,\nn\\
\wtd D\, \Big[ M^{-1}_{\gamma\a}\, M^{-1}_{\delta\beta} \, {*\wtd
F_\2^{\gamma\delta}}\Big]&=&
-2\td g\, M^{-1}_{\gamma[\a}\, 
{*\wtd D M_{\beta]\gamma}} - G_\3^\a\wedge G_\3^\beta +
\Phi\, M^{-1}_{\a\gamma}\, G_\1^\beta\wedge {*G_\2^\gamma} \label{s3ym}\\
&&\!\!\!\!\!\!\!\!\!\!\!\!\!\!\!\!
-\Phi\, M^{-1}_{\b\gamma}\, G_\1^\a\wedge {*G_\2^\gamma}
 -\ep_{\a\beta\gamma\delta}\, M_{\delta\lambda}\, G_\2^\gamma\wedge
{*G_\3^\lambda} - \ft12 \Phi^{-1}\, \ep_{\a\beta\gamma\delta}\, \wtd
F_\2^{\gamma\delta} \wedge {* \wtd S_\3^0}\,.\nn
\eea

    Finally, the scalar field equations (\ref{scalarsv}) give the
following:
\bea
d(\Phi^{-1}\, {*d\Phi}) &=& \Phi\, M_{\a\beta}\, {*G_1^\a}\wedge
G_\1^\beta + \Phi\, M^{-1}_{\a\beta}\, {*G_\2^\a}\wedge G_\2^\beta 
 -\Phi^{-1}\, {*\wtd S_\3^0}\wedge \wtd S_\3^0 +\ft15 Q\,,\nn\\
\wtd D(\Phi\, M_{\a\beta}\, {*G_\1^\beta}) &=&\Phi\,
M^{-1}_{\beta\gamma}\, {*G_\2^\gamma}\wedge \wtd F_\2^{\a\beta} 
- M_{\a\beta}\, {*G_\3^\beta}\wedge \wtd S_\3^0\,,\label{s3scalar}\\
\wtd D(M^{-1}_{\a\gamma}\, {*\wtd D M_{\gamma\beta}}) &=& \Phi\,
M_{\beta\gamma} {*G_\1^\gamma}\wedge G_\1^\a + M_{\beta\gamma}\,
{*G_\3^\gamma}\wedge G_\3^\a - \Phi\, M^{-1}_{\a\gamma}\,
{*G_\2^\gamma} \wedge G_\2^\beta \nn\\
&&\!\!\!\!\!
+ M^{-1}_{\a\gamma}\, M^{-1}_{\lambda\delta}\, {*\wtd
F_\2^{\gamma\delta}} \wedge \wtd F_\2^{\lambda\beta} +
2\td g^2( 2M_{\a\gamma}\, M_{\gamma\beta} - M_{\gamma\gamma}\,
M_{\a\beta})\, \ep_\7 - \ft15 \delta_{\a\beta}\, Q\,.\nn
\eea
In these equations, the quantity $Q$ is the limit of the trace term
multiplying $\delta_{ij}$ in (\ref{scalarsv}), and is given by
\bea
 Q&=&2\td g^2 \, \Big(2 M_{\a\beta}\, M_{\a\beta} -
(M_{\a\a})^2\Big)\, \ep_\7 -
M^{-1}_{\a\gamma}\, M^{-1}_{\beta\delta}\, {*\wtd
F_\2^{\a\beta}}\wedge \wtd F_\2^{\gamma\delta} \nn\\
&&+ \Phi^{-1}\, {*\wtd
S_\3^0}\wedge \wtd S_\3^0
 -2 \Phi\, M^{-1}_{\a\beta}\, {*G_\2^{\a}}
\wedge G_\2^{\beta} +
M_{\a\beta}\, {*G_\3^\a}\wedge
G_\3^\beta\,.\label{newtraceterm}
\eea

   Having obtained the seven-dimensional equations of motion for the
$SO(4)$-gauged limit, we can now seek a Lagrangian from which they can
be generated.  A crucial point is that the equations involving $\wtd
H_\4^0$ in (\ref{h0ha}) and (\ref{s3firstorder}) give
\be
d\wtd S_\3^0 =  \ft18 \ep_{\a_1\cdots \a_4}\, \wtd F_\2^{\a_1\a_2}\wedge 
                     \wtd F_\2^{\a_3\a_4}\,,
\ee
which allows us to strip off the exterior derivative by writing
\be
\wtd S_\3^0 = dA_\2 + \omega_\3\,,
\ee
where $\wtd S_\3^0$ is now viewed as a field strength with 2-form 
potential $A_\2$, and
\be
\omega_\3 \equiv  
\ft18 \ep_{\a_1\cdots \a_4}\, (\wtd F_\2^{\a_1\a_2}\wedge \wtd
A_\1^{\a_3\a_4} - \ft13 \td g\, \wtd A_\1^{\a_1\a_2}\wedge \wtd
A_\1^{\a_3\beta}\wedge \wtd A_\1^{\beta\a_4})\,.
\ee
We can now see that the equations of motion can be derived from the
following seven-dimensional Lagrangian, in which $A_\2$, and not its
field strength $\wtd S_\3^0\equiv dA_\2 +\omega_\3$, 
is viewed as a fundamental field:
\bea
{\cal L}_7 &=& R\, {*\oneone} - \ft{5}{16}\, \Phi^{-2}\,
{*d\Phi}\wedge d\Phi 
- \ft14 M^{-1}_{\a\beta}\, {*\wtd
D}M_{\beta\gamma}\wedge M^{-1}_{\gamma\delta}\, \wtd D M_{\delta\a}
-\ft12 \Phi^{-1}\, {*\wtd S_\3^0}\wedge \wtd S_\3^0\nn\\
&& -\ft14 M^{-1}_{\a\gamma}\, M^{-1}_{\beta\delta}\, {*\wtd
F_\2^{\a\beta}}\wedge \wtd F_\2^{\gamma\delta} - \ft12 \Phi\,
M^{-1}_{\a\beta}\, {*G_\2^\a}\wedge G_\2^\beta -\ft12\Phi\,
M_{\a\beta}\, {*G_\1^\a}\wedge G_\1^\beta \nn\\
&&- \ft12 M_{\a\beta}\,{*G_\3^\a}\wedge G_\3^\beta -\wtd V\,
{*\oneone}  
+\fft1{2\td g} 
\wtd D \wtd S_\3^\a\wedge \wtd S_\3^\a  
+\wtd S_\3^\a\wedge \wtd S_\3^0\wedge A_\1^{0\a}\label{d7lag0}\\
&&+\fft1{2\td g}\, 
\ep_{\a\beta\gamma\delta}\, \wtd S_\3^\a \wedge 
\wtd F_\2^{0\beta}\wedge \wtd
F_\2^{\gamma\delta} + \ft14 \ep_{\a_1\cdots\a_4}\, \wtd S_\3^0\wedge
\wtd F_\2^{\a_1\a_2}\wedge \wtd A_\1^{0\a_3}\wedge \wtd A_\1^{0\a_4}
+ \fft1{\td g}\wtd \Omega_\7\,,\nn
\eea
where $\wtd \Omega_\7$ is built purely from $\wtd A_\1^{\a\beta}$ and
$\wtd A_\1^{0\a}$.  It is defined by the requirement that its
variations with respect to $\wtd A_\1^{\a\beta}$ and $\wtd A_\1^{0\a}$
should produce the necessary terms in the equations of motion for these
fields.  Since it has a rather complicated structure, we shall not
present it here.  Note that the fields that are treated as
fundamental in this Lagrangian are the metric and the scalars
$(\Phi,M_{\a\beta})$, together with $(\chi_\a,\wtd A_\1^{\a\beta},
\wtd A_\1^{0\a}, \wtd S_\3^\a, A_\2)$, but it should be borne in mind
that $\Phi$ is not independent of $M_{\a\beta}$, because of the 
relation (\ref{phim}).  It can be useful, therefore, to define the
unimodular matrix $\wtd M_{\a\beta} \equiv \Phi^{-1/4}\, M_{\a\beta}$,
so that $\wtd M_{\a\beta}$ and $\Phi$ are independent fields.  The
scalar part of the Lagrangian (\ref{d7lag0}) then becomes
\be
{\cal L}_{\rm scal} = -\ft14 \Phi^{-2}\, {*d\Phi}\wedge d\Phi - \ft14 
\wtd M^{-1}_{\a\beta}\, {*\wtd
D}\wtd M_{\beta\gamma}\wedge \wtd M^{-1}_{\gamma\delta}\, 
\wtd D \wtd M_{\delta\a}\,.
\ee

\section{$\R\times S^3$ limit of the $S^4$ reduction}

   In the previous section, we obtained a scaling limit of the gauged
$SO(5)$ theory in seven dimensions, in which an $SO(4)$ gauging
survives.  In this section, we apply this scaling procedure to the
$S^4$ reduction Ansatz of section 2, showing that it leads to a
degeneration in which the 4-sphere becomes $\R\times S^3$.  We can
then reinterpret the reduction from $D=11$ as an initial ``ordinary''
Kaluza-Klein reduction step from $D=11$ to give the type IIA
supergravity in $D=10$, followed by a non-trivial reduction of the
type IIA theory on $S^3$, in which the entire $SO(4)$ isometry group
is gauged.\footnote{The $S^3$ reduction of type IIA supergravity
discussed in \cite{chamsab}, giving a seven-dimensional theory with
just an $SU(2)$ gauging, was rederived in \cite{nasvam} as a singular
limit of the $S^4$ reduction of $D=11$ supergravity that was obtained
in \cite{vann2}.  Since the $S^3$ reduction in \cite{chamsab} retains
only the left-acting $SU(2)$ of the $SO(4)\sim SU(2)_L\times SU(2)_R$
of gauge fields, the consistency of that reduction is guaranteed by
group-theoretic arguments, based on the fact that all the retained
fields are singlets under the right-acting $SU(2)_R$.  The subtleties
of the consistency of the $S^4$ reduction in \cite{vann2} are
therefore lost in the singular limit to $\R\times S^3$ discussed in
\cite{nasvam}, since a truncation to the $SU(2)_L$ subgroup of the
$SO(4)$ gauge group is made.  By contrast, the $\R\times S^3$ singular
limit that we consider here retains all the fields of the $S^4$
reduction in \cite{vann2}, and the proof of the consistency of the
resulting $S^3$ reduction of the type IIA theory follows from the
non-trivial consistency of the reduction in \cite{vann2}, and has no
simple group-theoretic explanation.}

\subsection{The $\R\times S^3$ reduction Ansatz}

    To take this limit, we combine the scalings of seven-dimensional
quantities derived in the previous section with an
appropriately-matched rescaling of the coordinates $\mu^i$ defined on
the internal 4-sphere.  As in \cite{jim}, we see that after splitting
the $\mu^i$ into $\mu^0$ and $\mu^\a$, these additional scalings
should take the form
\be
\mu^0 = \lambda^5\, \td\mu^0\,,\qquad \mu^\a =\td\mu^\a\,.\label{mulim}
\ee
In the limit where $\lambda$ goes to zero, we see that the original
constraint $\mu^i\, \mu^i=1$ becomes
\be
\td\mu^\a\, \td\mu^\a=1\,,
\ee
implying that the $\td\mu^\a$ coordinates define a 3-sphere, while the
coordinate $\td\mu^0$ is now unconstrained and ranges over the
real line $\R$.

    Combining this with the rescalings of the previous section, we
find that the $S^4$ metric reduction Ansatz (\ref{metel}) becomes
\be
d\hat s_{11}^2 = \lambda^{-2/3}\, \Big[ \wtd \Delta^{1/3}\, ds_7^2 +
\fft1{\td g^2}\, \wtd\Delta^{-2/3}\, M^{-1}_{\a\b}\, \wtd D\td\mu^\a
\, \wtd D\td\mu^\b+
\fft1{\td g^2}\, \wtd\Delta^{-2/3}\, \Phi\, (d\td\mu_0 + \td g\, \wtd
A_\1^{0\a}\, \td \mu^\a + \chi_\a\, \wtd D\td\mu^\a)^2 \Big]\,,
\label{r1s3met}
\ee
where
\be
\wtd\Delta \equiv M_{\a\beta}\, \td\mu^\a\, \td\mu^\beta\,.
\ee
Thus $\td\mu_0$ can be interpreted as the ``extra''
coordinate of a standard type of Kaluza-Klein reduction from $D=11$ to
$D=10$, with 
\be
d\hat s_{11}^2 = e^{-\fft16\phi}\, ds_{10}^2 + e^{\fft43 \phi}\,
(d\td\mu_0 + \cA_\1)^2\,.\label{1step}
\ee

    By comparing (\ref{1step}) with (\ref{r1s3met}), we can read off
the $S^3$ reduction Ansatz for the ten-dimensional fields.  Thus we
find that the ten-dimensional metric is reduced according to
\be
ds_{10}^2 = \Phi^{1/8}\, \Big[ \wtd\Delta^{1/4}\, ds_7^2 +\fft1{\td
g^2}\, \wtd\Delta^{-3/4}\,   M^{-1}_{\a\b}\, \wtd D\td\mu^\a
\, \wtd D\td\mu^\b\Big]\,,
\ee
while the Ansatz for the dilaton $\phi$ of the ten-dimensional theory
is
\be
e^{2\phi} = \wtd\Delta^{-1}\, \Phi^{3/2}\,.
\ee
Finally, the reduction Ansatz for the ten-dimensional Kaluza-Klein
vector is
\be
\cA_\1 =  \td g\, \wtd
A_\1^{0\a}\, \td \mu^\a + \chi_\a\, \wtd D\td\mu^\a\,.\label{1formans}
\ee
These results for the $S^3$ reduction of the ten-dimensional metric
and dilaton agree precisely with the results obtained in \cite{s3red}.
(Note that the field $\Phi$ is called $Y$ there, and our $M_{\a\b}$ is
called $T_{ij}$ there.)   Note that the field strength
$\cF_\2=d\cA_\1$ following from (\ref{1formans}) has the simple
expression
\be
\cF_\2 = \td g\, G_\2^\a\, \td\mu^\a + G_\1^\a\wedge \wtd
D\td\mu^\a\,.
\ee

    So far, we have read off the reduction Ans\"atze for those fields
of ten-dimensional type IIA supergravity that come from the reduction
of the eleven-dimensional metric.  The remaining type IIA fields come
from the reduction of the eleven-dimensional 4-form.  Under the
standard Kaluza-Klein procedure, this reduces as follows:
\be
\hat F_\4 = F_\4 + F_\3\wedge (d\td\mu^0 + \cA_\1)\,.\label{4fs1}
\ee
By applying the $\lambda$-rescaling derived previously to the $S^4$
reduction Ansatz (\ref{4form}) for the eleven-dimensional 4-form, and
comparing with (\ref{4fs1}), we obtain the following expressions for
the $S^3$ reduction Ans\"atze for the ten-dimensional 4-form and
3-form fields:
\bea
F_\4 &=&
\fr{\wtd{\D}^{-1}}{\td{g}^3}\,M_{\a\b}\,G_\1^\a\,\td{\m}^\b\wedge
\wtd{W} + \fr{\wtd{\D}^{-1}}{2\td{g}^2}\, \e_{\a_1\ldots\a_4}\, 
M_{\a_4\b}\td{\m}^\b\,G_\2^{\a_1}\wedge\wtd{D}\td{\m}^{\a_2}
\wedge\wtd{D}\td{\m}^{\a_3}\nn\\
& & - M_{\a\b}*G_\3^{\a}\td{\m}^{\b}
+ \fr1{\td{g}}\, G_\3^\a\wedge\wtd{D}\td{\m}^\a\,,\\
F_\3 &=& -\fr{\wtd{U}\wtd{\D}^{-2}}{\td{g}^3}\, \wtd W + 
\fr{\wtd\D^{-2}}{2\td{g}^3}\, \e_{\a_1\ldots\a_4}\,
M_{\a_1\b}\td{\m}^\b\, \wtd{D}M_{\a_2\g}\td{\m}^\g
\wedge\wtd{D}\td{\m}^{\a_3}\wedge\wtd{D}\td{\m}^{\a_4}\nn\\
& & +\fr{\wtd{\D}^{-1}}{2\td{g}^2}\, \e_{\a_1\ldots\a_4}\, 
M_{\a_1\b}\td{\m}^\b\,\wtd{F}_\2^{\a_2\a_3}\wedge\wtd{D}\td{\m}^{\a_4}
+ \fr1{\td g}\, \wtd{S}_\3^0\,,
\eea
where
\be
\wtd W \equiv \ft1{6}\, \ep_{\a_1\cdots \a_4}\, \td\mu^{\a_1}\, \wtd
D\td\mu^{\a_2} \wedge \wtd D\td\mu^{\a_3}\wedge \wtd D\td\mu^{\a_4}\,.
\ee

   It is also useful to present the expressions for the
ten-dimensional Hodge duals of the field strengths:
\bea
{\rm e}^{\fft12\f}\, {\bar *F_\4} &=& \fr1{\td{g}}\,\Phi\,
M_{\a\b}\, {*G_\1^\a\td{\m}^\b} - \fr1{\td{g}^2}\, \Phi \,M_{\a\b}^{-1}
\, {*G_\2^\a}\wedge\wtd{D}\td{\m}^\b + 
\fr{\wtd{\D}^{-1}}{\td{g}^4}\,M_{\a\b} G_\3^\a \td{\m}^\b\wedge
\wtd{W}\nn\\
& &+ \fr{\wtd{\D}^{-1}}{2\td{g}^3}
\e_{\a_1\ldots\a_4}\,M_{\a_1\b}\td{\m}^\b\,
M_{\a_2\g}\,
{*G_\3^\g}\wedge\wtd{D}\td{\m}^{\a_3}\wedge\wtd{D}\td{\m}^{\a_4}\, ,\\
{\rm e}^{-\f}\, {\bar *F_\3} &=& - \td{g}\wtd{U}\e_\7 - \fr1{\td{g}^3}
\Phi^{-1}\, {*\wtd{S}^0_\3}\wedge\wtd{W} \nn\\
& & + \fr1{2\td{g}^2} M_{\a\g}^{-1}\, M_{\b\d}^{-1}\,
{*\wtd{F}^{\a\b}_\2}\wedge\wtd{D}\td{\m}^\g\wedge\wtd{D}\td{\m}^\d
- \fr1{\td{g}}M_{\a\b}^{-1}\, {*\wtd{D}M_{\a\g}}
\td{\m}^\g\wedge\wtd{D}\,\td{\m}^{\b}\,,\nn\\
 e^{\fr32 \phi}\bar{*}{\cal F}_\2 &=&
\frac{\tilde{\Delta}^{-1}\Phi}{\tilde{g}^5}\, {*G_\2^\alpha}
\tilde{\mu}^\alpha\wedge \tilde{W} +
\frac{\tilde{\Delta}^{-1}\Phi}{2\tilde{g}^4} \,\epsilon_{\alpha_1\cdots
\alpha_4} \, M_{\alpha_1\beta}\, \tilde{\mu}^\beta
M_{\alpha_2\gamma}\, {*G_\1^\gamma}\wedge
\tilde{D}\tilde{\mu}^{\alpha_3}\wedge\tilde{D}\tilde{\mu}^{\alpha_4} \,.\nn
\eea
(Here we are using $\bar *$ to denote a Hodge dualisation in the
ten-dimensional metric $ds_{10}^2$, to distinguish it from $*$ which
denotes the seven-dimensional Hodge dual in the metric $ds_7^2$. )

\subsection{Verification of the reduction Ansatz}

     The consistency of the $S^3$ reduction of the type IIA theory
using the Ansatz that we obtained in the previous subsection is
guaranteed by virtue of the consistency of the $S^4$ reduction from
$D=11$.  It is still useful, however, to examine the reduction
directly, by substituting the Ansatz into the equations of motion of
type IIA supergravity.  By this means we can obtain an explicit
verification of the validity of the limiting procedures that we
applied in obtaining the $S^3$ reduction Ansatz.

   The bosonic Lagrangian for type IIA supergravity can be written as
\bea
{\cal L}_{10} &=& R\, {\bar *\oneone} - \ft12 {\bar *d\f}\wedge d\f -
\ft12 {\rm e}^{\fft32\f}\, {\bar *{\cal F}_\2}\wedge {\cal F}_\2 -
\ft12 {\rm e}^{\fft12\f}\, {\bar *F_\4}\wedge F_\4\nn\\
& & -\ft12 {\rm e}^{-\f}\, {\bar *F_\3}\wedge F_\3 + \ft12
dA_\3\wedge dA_\3\wedge A_\2\,,
\eea
where 
\be
F_\4 = dA_\3 - dA_\2 \wedge \cA_\1\,,\qquad
F_\3 = dA_\2\,,\qquad 
\cF_\2 = d\cA_\1\,.
\ee
(In this subsection, we use a bar where necessary to indicate
ten-dimensional quantities.)

    The equations of motion derived from the above Lagrangian are
\bea
d{\bar *d{\f}} &=& \ft12 {e}^{-{\f}}\, {\bar *F_\3}\wedge
F_\3 - \ft34 {e}^{\fft32\f}\, {\bar *{\cal F}_\2}\wedge
{\cal F}_\2 - \ft14 {\rm e}^{\fft12\f}\,{\bar *F_\4}\wedge
F_\4 \, ,\nn\\
d({e}^{\fr12{\f}}\,{\bar *F_\4}) &=& {F}_\4\wedge
{F}_\3 \, ,\nn\\
d({e}^{\fr32{\f}}\,{\bar *{\cal F}_\2}) &=& - 
e^{\fr12{\f}}\, {\bar *F_\4}\wedge
{F}_\3 \, ,\nn\\
d({e}^{-{\f}}\,{\bar *F_\3}) &=& \ft12 {F}_\4\wedge
{F}_\4 - {e}^{\fr12{\f}}\, {\bar *F_\4}\wedge
{\cal F}_\2\, .
\eea
Note that it is consistent to truncate the theory to the NS-NS sector,
namely the subsector comprising the metric, the dilaton and the 3-form
field strength.  This implies that it is possible also to perform an
$S^3$ reduction of the NS-NS sector alone, which was indeed
demonstrated in \cite{s3red}.  On the other hand it is not consistent
to truncate the theory to a subsector comprising only the metric, the
dilaton and the 4-form field strength, which again is in agreement
with the conclusion in \cite{s3red} that it is not consistent to
perform an $S^4$ reduction on such a subsector.  However, as we show
in section 5, there {\it is} a consistent $S^4$ reduction if we
include all the fields of the type IIA theory.
 
   The reduction Ansatz obtained in section (4.1) can now be
substituted into the type IIA equations of motion, to verify that it
indeed leads to the equations of motion for the $SO(4)$-gauged
seven-dimensional theory constructed in section 3.

\section{$S^4$ reduction of type IIA supergravity}

    We can also derive the Ansatz for the consistent $S^4$ reduction
of type IIA supergravity from the $S^4$ reduction Ansatz of
eleven-dimensional supergravity.  In this case we do not need to take
any singular limit of the internal 4-sphere, but rather, we extract
the ``extra'' coordinate from the seven-dimensional spacetime of the
original eleven-dimensional supergravity reduction Ansatz.  The
resulting six-dimensional $SO(5)$ gauged maximal supergravity can be
obtained from the Kaluza-Klein reduction of seven-dimensional gauged
maximal supergravity on a circle.

    We begin, therefore, by making a standard $S^1$ Kaluza-Klein
reduction of the seven-dimensional metric:
\be
ds_7^2 = e^{-2\a\varphi}\, ds_6^2 + e^{8\a\varphi}\, 
(dz+ \bar\cA_\1)^2\,,\label{sevensix}
\ee
where $\a=1/\sqrt{40}$.  With this parameterisation the metric
reduction preserves the Einstein frame, and the dilatonic scalar
$\varphi$ has the canonical normalisation for its kinetic term in six
dimensions.\footnote{We use a bar to denote six-dimensional fields, in
cases where this is necessary to avoid an ambiguity.}  Substituting
(\ref{sevensix}) into the original metric reduction Ansatz
(\ref{metel}), we obtain
\be
d\hat s_{11}^2 = \Delta^{1/3}\, e^{-2\a\varphi}\, ds_6^2 +
\fft1{g^2}\, \Delta^{-2/3}\, T_{ij}^{-1}\, D\mu^i\, D\mu^j +
\Delta^{1/3}\, e^{8\a\varphi}\, (dz+ \bar\cA_\1)^2\,.\label{firstgo}
\ee

    In order to extract the Ansatz for the $S^4$ reduction of type IIA
supergravity, we must first rewrite (\ref{firstgo}) in the form
\be
d\hat s_{11}^2 = e^{-\fft16 \phi}\, ds_{10}^2 + e^{\fft43\phi}\,
(dz + \cA_\1)^2\,,\label{delten}
\ee
which is a canonical $S^1$ reduction from $D=11$ to $D=10$.  It is not
immediately obvious that this can easily be done, since the
Yang-Mills fields $A_\1^{ij}$ appearing in the covariant differentials
$D\mu^i$ in (\ref{firstgo}) must themselves be reduced according to
standard Kaluza-Klein rules,
\be
A_\1^{ij} = \bar A_\1^{ij} + \chi^{ij}\, (dz+\bar\cA_\1)\,,
\ee
where $\bar A_\1^{ij}$ are the $SO(5)$ gauge potentials in six
dimensions, and $\chi^{ij}$ are six-dimensional axions.  Thus we have
\be
D\mu^i = \bD\mu^i + g\, \chi^{ij}\, \mu^j\, (dz+\bar\cA_\1)\,,
\ee
where 
\be
\bD\mu^i \equiv  d\mu^i + g\, \bar A_\1^{ij}\, \mu^j\,.
\ee
This means that the differential $dz$ actually appears in a much more
complicated way in (\ref{firstgo}) than is apparent at first sight.
Nonetheless, we find that one can in fact ``miraculously'' complete
the square, and thereby rewrite (\ref{firstgo}) in the form of
(\ref{delten}).

    To present the result, it is useful to make the following
definitions:
\bea
\Omega &\equiv & \Delta^{1/3}\, e^{8\a\varphi} + \Delta^{-2/3}\,
T_{ij}^{-1}\, \chi^{ik}\, \chi^{j\ell}\, \mu^k\, \mu^\ell\,,\nn\\
Z_{ij} &\equiv & T_{ij}^{-1} - \Omega^{-1}\, \Delta^{-2/3}\,
T_{ik}^{-1}\, T_{j\ell}^{-1}\, \chi^{km}\, \chi^{\ell n}\, \mu^m\,
\mu^n\,,\label{omsdef}
\eea
In terms of these, we find after some algebra that we can rewrite
(\ref{firstgo}) as
\be
d\hat s_{11}^2 =\Delta^{1/3}\, e^{-2\a\varphi}\, ds_6^2 +
\fft1{g^2}\, \Delta^{-2/3}\, Z_{ij}\, 
\bD\mu^i\, \bD\mu^j
+\Omega\, (dz+\cA_\1)^2\,,\label{second}
\ee
where the ten-dimensional potential $\cA_\1$ is given in terms of
six-dimensional fields by
\be
\cA_\1 = \bar \cA_\1 + \fft1{g}\, \Omega^{-1}\, \Delta^{-2/3}\,
T_{ij}^{-1}\, \chi^{jk}\, \mu^k\, \bar D\mu^i\,.\label{d101form}
\ee
This is therefore the Kaluza-Klein $S^4$ reduction Ansatz for the
1-form $\cA_\1$ of the type IIA theory.  Comparing (\ref{second}) with
(\ref{delten}), we see that the Kaluza-Klein reduction Ans\"atze for
the metric $ds_{10}^2$ and dilaton $\phi$ of the type IIA theory are
given by
\bea
ds_{10}^2 &=& \Omega^{1/8}\, \Delta^{1/3}\, e^{-2\a\varphi}\, 
ds_6^2  + \fft1{g^2}\, \Omega^{1/8}\, \Delta^{-2/3}\, Z_{ij}\, 
\bD\mu^i\,\bD\mu^j \,,\nn\\
e^{\fft43\phi} &=& \Omega\,.\label{d10metphi}
\eea

    The $S^4$ reduction Ansatz for the R-R 4-form $F_\4$ of the type
IIA theory is obtained in a similar manner, by first implementing a
standard $S^1$ Kaluza-Klein reduction on the various seven-dimensional
fields appearing in the $S^4$ reduction Ansatz (\ref{4form}) for the
eleven-dimensional 4-form $\hat F_\4$, and then matching this to a
standard $S^1$ reduction of $\hat F_\4$ from $D=11$ to $D=10$:
\be
\hat F_\4 = F_\4 + F_\3\wedge (dz+\cA_\1)\,.\label{f4f3}
\ee
Note that in doing this, it is appropriate to treat the 3-form fields
$S_\3^i$ of the seven-dimensional theory as field strengths for the
purpose of the $S^1$ reduction to $D=6$, {\it viz}.
\be
S_\3^i =  \bar S_\3^i + \bar S_\2^i\wedge (dz+ \bar\cA_\1)\,.
\ee
It is worth noting also that this implies that the reduction of the
seven-dimensional Hodge duals ${*S_\3^i}$ will be given by
\be
{*S_\3^i} = e^{4\a\varphi}\, {\bar * \bar S_\3^i}\wedge (dz+ \bar
\cA_\1) + e^{-6\a\varphi}\, {\bar *\bar S_\2^i}\,,
\ee
where $\bar *$ denotes a Hodge dualisation in the six-dimensional
metric $ds_6^2$. 

    With these preliminaries, it is now a mechanical, albeit somewhat
uninspiring, exercise to make the necessary substitutions into
(\ref{4form}), and, by comparing with (\ref{f4f3}), read off the
expressions for $F_\4$ and $F_\3$.  These give the Kaluza-Klein $S^4$
reductions Ans\"atze for the 4-form and 3-form field strengths of type
IIA supergravity.  We shall not present the results explicitly here,
since they are rather complicated, and are easily written down ``by
inspection'' if required.  For these purposes, the following
identities are useful:
\bea
(dz+\bar\cA_\1) &=& (dz+\cA_\1) - \fft1{g}\, \Omega^{-1}\, \Delta^{-2/3}\,
T_{ij}^{-1}\, \chi^{jk}\, \mu^k\, \bar D\mu^i\,,\nn\\
D\mu^i &=& T_{ij}\, Z_{jk}\, \bar D\mu^k + g\, \chi^{ij}\, \mu^j\,
(dz+\cA_\1)\,,\\
D X_i &=& \bar D X_i - \Omega^{-1}\, \Delta^{-2/3}\, \chi^{ij}\, X_j\, 
T_{k\ell}^{-1}\, \chi^{\ell m}\, \mu^m\, \bar D\mu^k + g\, \chi^{ij}\,
(dz+\cA_\1)\,,\nn
\eea
where in the last line $X_i$ represents any six-dimensional field in
the vector representation of $SO(5)$, and the covariant derivative
generalises to higher-rank $SO(5)$ tensors in the obvious way.

   If we substitute the $S^4$ reduction Ans\"atze given for the
ten-dimensional dilaton, metric and 1-form in (\ref{d10metphi}),
and (\ref{d101form}), together with those for
$F_\4$ and $F_\3$ as described above, into the equations of motion of
type IIA supergravity, we shall obtain a consistent reduction to six
dimensions.  This six-dimensional theory will be precisely the one
that follows by performing an ordinary $S^1$ Kaluza-Klein reduction on
the $SO(5)$-gauged maximal supergravity in $D=7$, whose bosonic
Lagrangian is given in (\ref{d7lag0}).

    It is perhaps worth remarking that the expression
(\ref{d10metphi}) for the Kaluza-Klein $S^4$ reduction of the type
IIA supergravity metric illustrates a point that has been observed
previously (for example in \cite{d4gauge,clpd45}), namely that the
Ansatz becomes much more complicated when axions or pseudoscalars are
involved.  Although the axions $\chi^{ij}$ would not be seen in the
metric Ansatz in a linearised analysis, they make an appearance in a
rather complicated way in the full non-linear Ansatz that we have
obtained here, for example in the quantities $\Omega$ and $Z_{ij}$
defined in (\ref{omsdef}).  They will also, of course, appear in the
Ans\"atze for the $F_\4$ and $F_\3$ field strengths.  It may be that
the results we are finding here could be useful in other contexts, for
providing clues as to how the axionic scalars should appear in the
Kaluza-Klein reduction Ansatz.

\section{Conclusions}

     In this paper, we have obtained a consistent 3-sphere reduction
of type IIA supergravity, in which all the massless $SO(4)$ gauge
bosons associated with the isometry group of the 3-sphere are
retained.  The resulting seven-dimensional gauged supergravity will,
accordingly, be maximally supersymmetric.  It is, however, not a
theory that admits an AdS$_7$ vacuum solution, but rather, it allows a
domain wall as its ``most symmetric'' ground state.  Since the
3-sphere is isomorphic to $SU(2)$ our construction can be set in the
context of a string propagating in a group-manifold background.
However, the reduction of fields that we considered here goes beyond
what is customarily included in such cases, since we can retain the
entire set of $SO(4)\sim SU(2)_L\times SU(2)_R$ Yang-Mills fields, and
not merely those of either the left-acting or right-acting $SU(2)$. 

    It is perhaps worth emphasising that although we can interpret the
$\R\times S^3$ limit of the $S^4$ reduction from $D=11$ as an $S^3$
reduction of the type IIA theory, we cannot reverse the roles of the
$\R$ and $S^3$ factors and interpret the limit as an $S^3$ reduction
of eleven-dimensional supergravity to give an $SO(4)$-gauged
supergravity in $D=8$, which then undergoes a further reduction to
$D=7$.  The reason for this is that when the limiting procedure is
applied to the $\mu^i$ coordinates of $S^4$, as in (\ref{mulim}), the
original coordinate $\mu^0$ is set to zero, and so all fields
necessarily become independent of the rescaled coordinate $\td \mu^0$
on the $\R$ factor.  This means that the consistent reduction
involving $S^3$ in the limit works only if the fields are all assumed
to be independent of the coordinate $\td\mu^0$ as well, and so there
would be no possibility of extracting an eight-dimensional covariant
theory by just considering the $S^3$ factor in the $\R\times S^3$
reduction. 

    The consistent $S^3$ reduction of type IIA supergravity that we
have constructed in this paper represents another element in the
accumulating body of examples where ``remarkable'' Kaluza-Klein sphere
reductions exist, even though there is no known group-theoretic
explanation for their consistency.  What is still lacking is a deeper
understanding of why they should work.  One might be tempted to think
that supersymmetry could provide the key, but this evidently cannot in
general be the answer, since there are examples such as the consistent
$S^3$ and $S^{D-3}$ reductions of the $D$-dimensional low-energy limit
of the bosonic string (in arbitrary dimension $D$) \cite{s3red} which
are obviously unrelated to supersymmetry.  

       As we discussed in introduction, we expect further examples of
consistent sphere reduction in type IIA and type IIB supergravities.
In particular, for non-trivial vacuum NS-NS flux, we expect that it is
consistent to reduce both type IIA and type IIB on $S^3$ and $S^7$.
For non-trivial vacuum R-R flux, we expect that it is consistent to
reduce the type IIA theory on $S^n$ with $n=2,4,6,8$ and for the type
IIB theory on $S^n$ with $n=1,3,5,7$.  The resulting maximal gauged
supergravities in the lower dimensions in general have domain-walls
rather than AdS as vacuum solutions, except in the case $n=5$ for type
IIB.  We constructed two such examples in this paper, namely the $S^3$
and $S^4$ reductions of the type IIA theory.  These domain-wall
supergravities provide useful tools with which to explore the Domain
Wall/QFT correspondence.

\section*{Acknowledgements}

    C.N.P.~is grateful for hospitality at the University of
Pennsylvania during the early stages of this work, and at the
Caltech-USC Center for Theoretical Physics during its completion.

\end{document}